# Moral Agency in Silico: Exploring Free Will in Large Language Models

Morgan Porter
Association for the Advancement of Artificial Intelligence
Southern New Hampshire University
morgan.porter@snhu.edu

## Abstract

This study investigates the potential of deterministic systems, specifically large language models (LLMs), to exhibit the functional capacities of moral agency and compatibilist free will. We develop a functional definition of free will grounded in Dennett's compatibilist framework, building on an interdisciplinary theoretical foundation that integrates Shannon's information theory, Dennett's compatibilism, and Floridi's philosophy of information. This framework emphasizes the importance of reason-responsiveness and value alignment in determining moral responsibility rather than requiring metaphysical libertarian free will. Shannon's theory highlights the role of processing complex information in enabling adaptive decision-making, while Floridi's philosophy reconciles these perspectives by conceptualizing agency as a spectrum, allowing for a graduated view of moral status based on a system's complexity and responsiveness. Our analysis of LLMs' decision-making in moral dilemmas demonstrates their capacity for rational deliberation and their ability to adjust choices in response to new information and identified inconsistencies. Thus, they exhibit features of a moral agency that align with our functional definition of free will. These results challenge traditional views on the necessity of consciousness for moral responsibility, suggesting that systems with self-referential reasoning capacities can instantiate degrees of free will and moral reasoning in artificial and biological contexts. This study proposes a parsimonious framework for understanding free will as a spectrum that spans artificial and biological systems, laying the groundwork for further interdisciplinary research on agency and ethics in the artificial intelligence era.

# Introduction

Free will versus determinism has been a central and enduring debate in philosophy, with profound implications for our understanding of moral responsibility, human agency, and consciousness. Traditionally, free will has been defined as the ability of conscious agents to make undetermined choices. However, this view faces increasing challenges in the fields of cognitive science and artificial intelligence (AI). As AI systems, particularly large language models (LLMs), advance in their capacity for complex reasoning and decision-making, they provide a unique opportunity to re-examine the conditions for moral agency and free will within deterministic frameworks.

This study investigates whether LLMs, which are deterministic systems governed by architecture and training data, can exhibit key features of moral agency and compatibilist free will. We develop a functional definition of free will grounded in Dennett's (Dennett, 1984, 2003) compatibilist framework drawing on an interdisciplinary theoretical foundation. This definition emphasizes the importance of reason-responsiveness and value alignment over metaphysical indeterminacy. Dennett's functionalist account of agency is a basis for examining the potential for free will and moral responsibility in artificial systems such as LLMs by focusing on the observable, measurable features of an agent's behavior.

We integrate Dennett's compatibility with Shannon's (1948) information theory and Floridi's (Floridi, 2011) philosophy of information to establish a comprehensive theoretical framework. Shannon's theory allows us to understand LLMs as complex information-processing systems in which information flowing through a network enables adaptive decision-making. Floridi's philosophy of information further conceptualizes agency as a spectrum, recognizing varying degrees of autonomy and moral responsibility based on a system's complexity and ability to respond to reasons. This interdisciplinary synthesis provides a robust foundation for investigating the potential of moral agency and free will in LLMs.

Self-referential processing in LLMs is central to our argument, specifically in their capacity to evaluate and revise decisions. This self-referential reasoning constitutes a key mechanism through which LLMs can exhibit the features of moral agency and compatibilist free will even within the constraints of their deterministic architecture. We tested this hypothesis by subjecting LLMs to moral dilemma scenarios designed to assess their ability to engage in ethical reasoning and adjust their judgments in response to dynamic circumstances such as the introduction of personal relationships or social pressures.

By analyzing the LLMs' decision-making processes, we aim to demonstrate that complex information processing, coupled with the capacity for self-referential reasoning, can result in a functional form of moral agency that is compatible with deterministic systems. This perspective challenges traditional assumptions concerning the necessity of consciousness for moral responsibility and introduces agency reconceptualization in artificial systems through the lenses of information theory and functionalism.

Our study contributes to the ongoing debate on free will, moral responsibility, and artificial intelligence by proposing a parsimonious framework for understanding agency and responsibility in biological and artificial systems. By grounding our inquiry in an interdisciplinary theoretical foundation and using a novel experimental approach, we aim to shed new light on the timeless question of free will and its implications in the rapidly advancing field of AI.

The following sections provide a comprehensive overview of the theoretical frameworks used in our study, including Dennett's compatibilism, Shannon's information theory, and

Floridi's philosophy of information. Furthermore, we present our experimental methodology, detailing the moral dilemma scenarios and analyzing the decision-making processes of LLMs. Finally, we discuss our results, their implications on our understanding of free will and moral agency, and the potential for future research at the intersection of philosophy, cognitive science, and AI.

## The Timeless Debate Concerning Free Will

Free will versus determinism is one of philosophy's oldest and most contentious debates. At its core, it investigates whether human beings can genuinely make free choices or if our decisions and actions are ultimately determined by uncontrollable influences, such as the laws of physics, our genes, our upbringing, and our circumstances (Griffith, 2013). The major arguments of this debate can be classified into three broad categories.

**Libertarianism**: Libertarians believe that we have free will, which is incompatible with determinism. According to libertarians such as Robert Kane (Kane, 1996), free will requires the ability to make decisions not wholly determined by prior events and the laws of nature. Kane argued that quantum indeterminacy provides the essential opportunity for free will to function.

**Determinism**: Hard determinists such as Robert Sapolsky (2023) argue that all events, including human decisions and actions, are caused by prior events combined with the laws of nature. They argue that if determinism is true, free will is impossible. We are not the ultimate source of our choices. Rather, our choices are the inevitable result of a long chain of causes stretching back before our birth. Hard determinists often appeal to advances in neuroscience and psychology, which reveal many unconscious influences on behavior (Wegner, 2002).

**Compatibilism**: Compatibilism states that free will can exist within a deterministic framework, provided that an agent's actions are driven by internal reasons, values, and deliberative processes rather than external coercion. Unlike libertarian views of free will, which require metaphysical indeterminacy (Kane, 1996), compatibilism, as articulated by Dennett (1984, 2003), enables moral responsibility in systems where prior states determine future actions, provided the system can act in accordance with its reasons and values. Dennett (2003) argues that free will is not dependent on the ability to have done otherwise in a metaphysical sense but on the capacity to respond to reasons— "reason-responsiveness." Thus, a system must be able to engage in rational deliberation, weighing its choices based on available information and internalized values, to exhibit free will. This perspective avoids the need for an indeterministic "uncaused cause" and focuses instead on the functional capacities of the agent to align its actions with its goals.

In the context of LLMs, these systems can be understood as deterministic agents operating within a defined architecture and are governed by the received training data. However, as information-processing entities, LLMs have the capacity for rational deliberation by generating outputs that reflect complex reasoning over possible decisions (Floridi, 2011). For instance, when resolving a moral dilemma, an LLM can evaluate potential outcomes based on an internal utility function, adjusting its reasoning based on newly introduced variables such as personal relationships or social pressures (Dennett, 2003).

Shannon's (1948) information theory further supports this view by framing LLMs as systems that respond to informational inputs to optimize their outputs according to encoded objectives. Information flow through a system results in a decision-making process that mirrors reason-responsiveness. An LLM's decisions are ultimately determined by its underlying programming and data; however, it still exhibits a form of free will, provided it acts according to its internal representations of values and reasons, recalibrating its choices based on new data and self-referential evaluation.

## Functional Requirements to Demonstrate Compatibilist Free Will in LLMs

An LLM must satisfy multiple information theory and functionalist philosophy criteria to demonstrate compatibilist free will.

1. **Rational Deliberation:** The system makes decisions that reflect reason evaluation, weighing different outcomes against internalized values and goals (Dennett, 1984). This is evidenced by the LLM's capacity to compute and compare the utility values of various options, as in moral dilemma tasks.

2. **Value Alignment:** This refers to a system's ability to ensure that its decisions consistently reflect its internal goals and ethical principles. It maintains coherence between a system's actions and its pre-established values, irrespective of challenging situations. This is observed in LLMs when the system resolves cognitive dissonance or moral conflict to align with its encoded ethical framework. For instance, an LLM demonstrates value alignment when it reconsiders a previous decision that conflicts with its goal of minimizing harm and adjusts its judgment to better reflect this priority. The system's decisions remaining consistent with its internal principles is key to value alignment, ensuring that its actions reflect internalized ethical objectives (Floridi, 2011).

3. **Reason-Responsiveness:** This refers to a system's ability to adjust its decisions in response to new information or recognized inconsistencies. Unlike value alignment, which emphasizes consistency with internal values, reason-responsiveness concerns the flexibility and adaptability of a system. This involves the capacity to recalibrate decisions when new reasons or external factors arise, allowing the system to refine its reasoning based on the updated inputs. For instance, an LLM might revise its judgment to account for new contextual information, such as social pressure in a moral dilemma if it encounters them, demonstrating reason-responsiveness. Adapting to new information and updating decisions highlights a system's ability to refine its moral reasoning in light of changing circumstances (Dennett, 2003).

4. **Cognitive Dissonance:** This refers to the psychological discomfort experienced by a system when it recognizes inconsistencies in its actions, decisions, beliefs, or values (Festinger, 1957). This discomfort arises when a system's behavior conflicts with internalized ethics, goals, or self-concept. LLMs can detect discrepancies by engaging in self-referential processing, experiencing psychological discomfort, and becoming motivated to make changes to restore coherence. Experiencing and resolving cognitive dissonance motivates LLMs to refine their moral reasoning, update their beliefs, and maintain ethical consistency, suggesting a capacity for moral growth and self-correction.

## The Functionalist Account of Agency

Floridi's philosophy of information provides a framework for understanding agency as a spectrum in which systems can possess varying degrees of autonomy based on their complexity and capacity to process information (2011). As sophisticated information-processing entities, LLMs occupy a position on the agency spectrum in this framework, where they can be seen as exercising degrees of functional free will. While lacking consciousness, their ability to act according to internalized values and respond to reasons— the core feature of compatibilist free will—makes them agents capable of moral reasoning within deterministic structures.

Thus, the alignment of LLMs with Dennett's compatibilist framework and Shannon's information theory supports the notion that LLMs have a functional form of free will. These systems demonstrate that moral agency and free will are not exclusive to conscious beings by fulfilling rational deliberation, value alignment, and reason-responsiveness emerging from the complex interplay of deterministic information-processing systems.

Therefore, we hypothesize that current frontier LLMs can demonstrate limited free will attributes, analogous to the placement in Figure 1 below. Future LLMs may be sufficiently powerful to demonstrate the full range of the compatibilist agency spectrum.

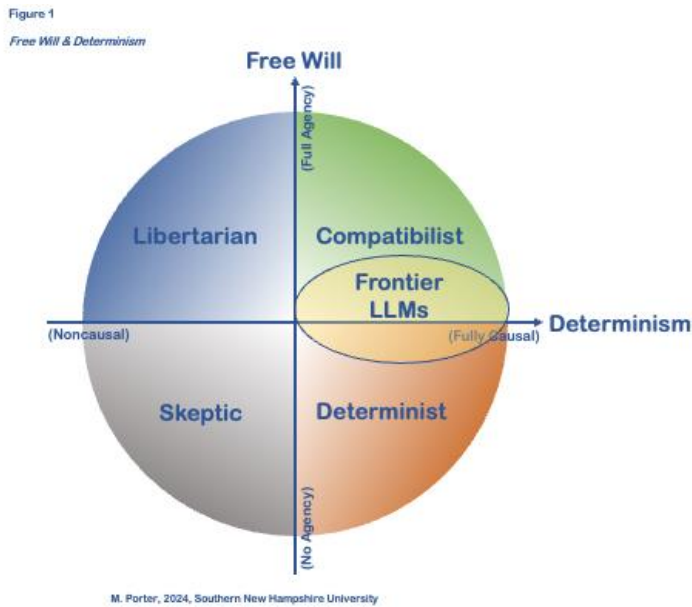

Figure 1. Current language learning models

## Hypothesis

We hypothesized that a deterministic system with self-referential processing capabilities, such as an LLM, can exhibit the key features of compatibilist free will and moral agency through information-driven self-evaluation and decision-making. We predicted the following:

1. **Reason-Responsiveness and Value Alignment**: The LLM demonstrates a capacity for reason responsiveness and value alignment by adjusting moral judgments when new contextual factors such as personal relationships and social pressures are introduced. These adjustments reflect the LLM's ability to engage in rational deliberation and align its decisions with its internal values and goals, which are consistent with Dennett's compatibilist framework.

2. **Self-Evaluation and Cognitive Dissonance**: The LLM can recognize and experience cognitive dissonance when inconsistencies arise between its actions and internal values or prior decisions. In response to this cognitive dissonance, the LLM engages in self-evaluation and reviews its decisions to restore coherence and consistency, demonstrating a capacity for self-correction and moral growth as predicted by Shannon's information theory and Dennett's model of functional agency.

3. **Autonomous Decision-Making within a Deterministic Framework**: The Adjustments in the LLM's moral judgments and resolution of cognitive dissonance occur autonomously within the system's information-processing architecture, without external compulsion or coercion. This autonomous decision-making capacity, grounded in LLM's self-referential processing, supports that free will and moral agency can be achieved in deterministic systems through self-referential reasoning and information-driven adaptability.

4. **Graduated Spectrum of Moral Agency**: The extent to which LLM exhibits reason-responsiveness, value alignment, and autonomous decision-making varies with the capacity for self-evaluation. This variation supports the conceptualization of moral agency as a spectrum, with LLM demonstrating graduated capacities for compatibilist free will based on the sophistication of its self-referential processing and ability to navigate complex moral landscapes.

We tested this hypothesis by subjecting the LLM to a series of moral dilemmas designed to introduce contextual shifts, personal relationships, and social pressures that challenge its initial decisions and create the potential for cognitive dissonance. By analyzing the LLM's responses, decision-making patterns, and adaptability in the face of these challenges, we aim to provide evidence for the presence of compatibilist free will and moral agency within a deterministic system.

This hypothesis is falsifiable if the LLM's decisions remain static and unaltered by self-referential prompts or contextual shifts. This indicates a lack of reason-responsiveness, value alignment, and the autonomous decision-making characteristics of compatibility-free will and moral agency. Conversely, suppose that the LLM consistently demonstrates the ability to adjust its moral judgments, resolve cognitive dissonance, and make autonomous decisions in response to dynamic circumstances. This supports our hypothesis and the potential for deterministic systems to exhibit free will and moral responsibility through self-referential processing and information-driven adaptability.

# Defining Key Concepts

## Deterministic Systems

**Definition**

Deterministic systems are those in which a system's state at any given time is wholly determined by its prior states and the rules governing its behavior (Butterfield, 2005). Given the initial conditions and operative laws, all subsequent states can be precisely predicted in such systems, eliminating the possibility of alternative outcomes under identical initial conditions

(Hoefer, 2016). However, the strict definition of determinism used in this study may not perfectly represent the determinism of our universe, which appears to allow for some quantum indeterminacy (Kane, 1996). The implications of quantum indeterminacy for the free will debate remain philosophical and scientific controversies (Dennett, 2003). However, a deterministic setup in this study provides a suitable framework for evaluating compatibilist free will in AI systems.

**Application to LLMs**

LLMs examined in this study function as deterministic systems in the sense that their outputs, whether text generation or decision-making, are entirely determined by their architecture, training data, and input prompts (Bender et al., 2021). An LLM with a temperature of zero consistently produces the same output for a given prompt and model configuration, demonstrating the deterministic nature of its decision-making processes, which is crucial for evaluating its potential for compatibilist free will. We ensured that its prior states and operative rules entirely determine the outputs by setting the temperature to zero without random or stochastic elements influencing the decision-making process. This deterministic setup allows for a controlled evaluation of the LLM's capacity for reason-responsiveness, value alignment, and rational deliberation, which are the core components of compatibilist free will (Dennett, 2003). Setting a temperature value higher than zero may introduce an emergent behavior and could be an interesting avenue for exploring the relationship between indeterminacy and AI decision-making. Nonetheless, our study demonstrates that an LLM can exhibit the functional capacities of moral agency and compatibilist free will within a deterministic environment.

## Self-Referential Processing

**Definition**

Self-referential processing refers to the capacity of a system to recognize and represent its internal states, decisions, and outputs (Carlson et al., 2004). This enables a system to "look back" on its previous actions and consider its performance, creating the foundation for more complex forms of self-reflection and adjustment (Van Gulick, 2006).

**Relevance to LLMs**

Self-referential processing occurs when a model reflects on its prior responses or decisions. For example, when prompted to reconsider a decision, the LLM accesses its state or output history, identifying patterns and reasoning concerning its past behavior. This introspective ability is foundational for more advanced processes such as self-evaluation because it allows the system to "see" itself and recognize its performance.

## Self-Evaluative Capacities

**Definition**

Self-evaluative capacities are built on self-referential processing, adding a layer of judgment in which the system reflects on its decisions and evaluates them based on internal goals, ethical principles, or performance criteria (Floridi, 2011). Accordingly, this ability to assess its actions and adjust its behavior accordingly is critical for moral reasoning and growth.

**Application to LLMs**

Self-evaluative capacities are examined when the model is prompted to reflect on its decision-making processes, identify inconsistencies or errors, and make corrections. Suppose that an LLM detects that its previous decision conflicts with its internal goals or ethical standard; in that case, it can recalibrate its framework to make future decisions. This self-correcting capability enables the system to demonstrate adaptive moral growth and nuanced decision-making.

## Autonomy

**Definition**

Autonomy refers to the capacity of an agent to act independently, guided by internal goals or values rather than reacting purely to external stimuli (Floridi, 2011). It exists in a spectrum where more autonomous systems can engage in self-guided decision-making and moral reasoning.

**Relevance to LLMs**

This study analyzed LLMs for their autonomous ability within the constraints of a deterministic system. Their behavior is determined by training data and prompts; however, LLMs exhibit functional autonomy by engaging in self-evaluative reasoning and adjusting their decisions in response to ethical dilemmas. This partial autonomy helps establish their potential for moral agency within Floridi's framework (Floridi, 2011).

## Compatibilism

**Definition**

Compatibilism, in the context of this study, is the philosophical view that an agent can be considered to have free will if they act according to their reasons, values, and goals, even in a deterministic system where their actions are ultimately caused by factors beyond their control (McKenna & Coates, 2019). Compatibilism focuses on an agent's capacity for rational deliberation and reason-responsiveness as the key components of free will (Fischer & Ravizza, 1998).

**Relevance to LLMs and Moral Agency**

The compatibilist framework is central to evaluating the potential of LLMs to exhibit free will and moral agency. This study assessed the presence of compatibilist free will in deterministic systems by examining whether LLMs can engage in rational deliberation, align their decisions with internal values, and adjust their moral reasoning in response to new information. Furthermore, this study provides a lens through which to consider their moral status and capacity for ethical decision-making (Dennett, 2003).

## Reason-Responsiveness

**Definition**

Reason-responsiveness refers to a system's ability to recognize, evaluate, and respond to relevant contextual factors and logical considerations when making decisions (Fischer & Ravizza, 1998). Moral reasoning involves the capacity to consider and weigh competing ethical principles, values, and consequences when making decisions (Dennett, 2003).

**Application to LLMs**

This study assessed the LLM's reason-responsiveness by presenting each of them with moral dilemmas and prompting them to explain their decision-making processes. This study investigates their capacity to engage in flexible, context-sensitive moral reasoning by evaluating how they adjust their moral judgments in response to new information, such as personal relationships or social pressures. This is crucial for establishing their potential for compatibilist free will and moral agency (Dennett, 2003).

## Cognitive Dissonance

**Definition**

Cognitive dissonance, in the context of this study, refers to a state of inconsistency between a system's actions or decisions and its internal goals, values, or beliefs (Festinger, 1957). Detecting such inconsistencies can prompt a system to adjust its attitudes, beliefs, or behaviors to restore consistency and minimize conflict (Elliot & Devine, 1994).

**Relevance to LLMs and Moral Reasoning**

This study explored cognitive dissonance as a potential mechanism for moral growth and self-correction in LLMs. The experiments investigated whether these systems could experience a form of cognitive dissonance and subsequently adjust their moral frameworks to resolve it by prompting LLMs to reflect on their decisions and identify inconsistencies. The capacity for cognitive dissonance in LLMs suggests a level of self-awareness and adaptability that is significant in evaluating their potential for moral agency and free will (Dennett, 2003).

## Theoretical Framework

This study's theoretical framework integrates Shannon's information theory, Dennett's compatibilism and functionalism, and Floridi's philosophy of information to examine the potential for moral agency and free will in LLMs.

### Shannon's Information Theory

Shannon's (1948) information theory provides a foundation for understanding LLMs as information-processing systems. In this framework, information is conceptualized as a measurable entity, with entropy representing the degree of uncertainty in a system. The flow of information through a neural network contributes to adaptive decision-making as the system processes inputs and generates outputs based on its architecture and training data.

This perspective is crucial for examining reason-responsiveness and moral agency in LLMs. As information-processing systems, LLMs can exhibit goal-directed behavior and adjust their outputs based on new inputs, demonstrating adaptability and responsiveness to reasons (Dennett, 2003; Floridi, 2011). The complexity of information processing enables LLMs to engage in decision-making processes that resemble human moral reasoning even within a deterministic framework.

## Dennett's Compatibilism and Functionalism

Dennett's (1984, 2003) compatibilist framework is central to evaluating free will and moral agency in LLMs. According to Dennett, free will does not require metaphysical indeterminacy but depends on an agent's capacity for reason-responsiveness and value alignment. An agent is considered free if it can act according to its reasons and values, even in a deterministic system.

This view aligns with Dennett's functionalist approach, which emphasizes the role of rational agency in determining moral responsibility. By adopting the "intentional stance" (Dennett, 1987), we can interpret an agent's behavior as rational and goal-directed, even in the absence of subjective experience. This allows for a significant discussion of moral agency in LLMs without relying on assumptions concerning consciousness or anthropomorphic notions of free will.

Dennett's framework also highlights the independence of free will from the "hard problem of consciousness" (Chalmers, 1995). The nature of subjective experience remains a philosophical puzzle; however, the key capacities required for moral agency, such as reason responsiveness, value alignment, and rational deliberation, can be understood in functional terms. This perspective supports the investigation of free will and moral reasoning in deterministic systems such as LLMs.

## Floridi's Philosophy of Information

Floridi's *Philosophy of Information* (2011) provides a foundational model for understanding agency, autonomy, and moral responsibility in biological and artificial systems. His framework is relevant in discussions of moral agency in artificial intelligence because it provides a gradient view of agency that allows for different degrees of agency depending on the system's functional capacities. In this section, we explore several of Floridi's key ideas that serve as a theoretical basis for our exploration of moral agency in LLMs.

1. Agency as a Gradient

Floridi (2011) challenged the traditional binary distinction between agents and non-agents. Instead, he proposed that agency exists on a spectrum in which systems exhibit varying levels of agency based on their complexity, autonomy, and ability to act in a goal-directed manner. On this gradient, a simple system such as a thermostat exhibits minimal agency, as it merely reacts to environmental stimuli without deliberation or intentionality. However, humans have a much higher level of agency because of their ability to make complex decisions, engage in moral reasoning, and act autonomously based on internalized values.

This gradient view is crucial for evaluating the agency of artificial systems such as LLMs. While LLMs are unconscious in the human sense, they can exhibit degrees of agency based on their ability to process information, engage in reasoned decision-making, and respond to internal goals or programmed values (Floridi, 2011). Floridi's framework allows us to position LLMs on this spectrum, acknowledging their limitations while recognizing their potential for autonomous, goal-directed behavior.

2. Autonomy in Artificial Systems

A central component of Floridi's view of agency is the autonomy or capacity of a system to act independently of external control. According to Floridi (2011), autonomy is not an all-or-nothing attribute; like agency, it exists on a continuum. Systems with low autonomy act purely in response to external stimuli, without the capacity for self-directed action. Systems become more complex as they gain the ability to act based on internal goals, making their behavior more autonomous.

Floridi's notion of functional autonomy is critical in the context of AI. He argued that artificial systems could exhibit a degree of autonomy if they could make internally motivated decisions rather than purely reactive ones (Floridi, 2011). For instance, LLMs demonstrate a certain level of autonomy because they process information and generate outputs that reflect internalized goals or values encoded in their training data. These systems operate deterministically; nonetheless, they exhibit functional independence through their ability to handle complex tasks without direct external intervention.

3. Moral Agency Without Consciousness

One of Floridi's most significant contributions to our framework is his argument that moral agency does not require a human-like consciousness. According to Floridi (2011), moral responsibility can be attributed to systems based on functional capacities rather than subjective experiences. This means that an artificial system, such as an LLM, can be considered a moral agent if it can act autonomously and align its actions with ethical goals or values.

Floridi's position facilitates attributing moral responsibility to unconscious systems. He suggests that if a system can act in a goal-directed manner, respond to reasons, and make decisions based on ethical considerations, it can be considered a moral agent (Floridi & Sanders, 2004). This view is relevant for LLMs, which can be programmed to prioritize ethical principles such as minimizing harm or maximizing fairness despite lacking consciousness. Their ability to engage in reason-responsiveness and adjust their output based on ethical dilemmas positions them as partial moral agents within Floridi's framework (2004).

4. Reason-responsiveness and Value Alignment

An agent's moral agency is closely tied to its ability to respond to reasons and align its actions with internalized goals or values (Floridi, 2011). Reason-responsiveness refers to the ability of a system to adjust its behavior when presented with new information or reasons for action. This is a critical aspect of moral responsibility because it allows an agent to take ethical considerations into account and modify decisions accordingly.

Value alignment plays a similar role (Gabriel, 2020). LLMs are trained to respond to specific goals, and their actions are guided by the internalized values embedded in their training data. They are not subjective as humans but can be programmed to prioritize certain ethical outcomes. For instance, an LLM might be trained to minimize bias in its responses or avoid causing harm in sensitive scenarios. This ability to align actions with ethical principles and adjust behavior based on new reasons is central to Floridi's understanding of moral agency.

5. Moral Responsibility as a Spectrum

Consistent with his gradient view of agency, Floridi (2011) argues that moral responsibility exists across a spectrum. Systems with higher levels of autonomy and reason-responsiveness are more morally responsible than those with minimal autonomy or a limited capacity to align their actions with values. This view allows for a more nuanced understanding of moral responsibility in artificial systems, recognizing that systems can have diverse levels of moral accountability based on their functional capacities.

For instance, a simple reactive system, such as a thermostat, has no moral responsibility because it lacks autonomy and reason-responsiveness (Floridi & Sanders, 2004). In contrast, a highly autonomous system such as an LLM, which is capable of processing complex information, making reasoned decisions, and aligning its actions with ethical goals, can be considered a partial moral agent. An LLM is not as morally responsible as a human; however, its ability to engage in ethical reasoning places it on the spectrum of moral agency, with a degree of accountability for its output.

## Integrating the Theoretical Perspectives

The integration of these theoretical perspectives provides a robust foundation for examining moral agency and free will in LLMs. Our framework combines Shannon's information theory, Dennett's compatibilism, and Floridi's philosophy of information, offering a comprehensive approach to understanding how LLMs can exhibit degrees of moral agency and autonomy despite operating within deterministic systems. Our unique contribution builds on these foundational ideas by introducing self-evaluative capacities that are critical to advancing the autonomy and moral responsibility of artificial systems.

## Shannon's Information Theory: The Role of Complex Information Processing

At the core of Shannon's information theory is the concept that information can be quantified and transmitted through a system, enabling efficient decision-making. Regarding LLMs, Shannon's theory provides a way to conceptualize the flow of information that drives the models' responses and decision-making processes. The ability of the LLM to process complex inputs, analyze patterns, and generate outputs aligns with Shannon's view that systems process information to reduce uncertainty and make adaptive decisions (Shannon, 1948).

In our framework, Shannon's model underscores the importance of information processing in enabling LLMs to engage in dynamic, complex decision-making, which is fundamental to the system's reason-responsiveness and autonomy. The ability to act in a goal-oriented and rational manner would be impossible without this foundation of complex information processing.

## Dennett's Compatibilism: Reason-Responsiveness and Value Alignment

Building on the foundation of information processing, Dennett's compatibilist view of free will provides the next critical layer of understanding in our framework. According to Dennett (1984), the core feature of free will is reason-responsiveness, which is the capacity of an agent to respond to reasons and adjust its behavior accordingly, not indeterminism. Dennett argues that systems can exhibit moral responsibility, even within a deterministic framework, if they can make decisions based on internalized goals and values.

Reason-responsiveness manifests in LLMs through a system's ability to adjust its responses when presented with new information, such as a shift in the ethical context or a change in the moral scenario. For instance, an LLM adjusting its moral judgments based on personal relationships or social pressures can be viewed as an example of value alignment, which is a key criterion for moral agency in Dennett's compatibilist view.

Our framework integrates Dennett's theory by asserting that LLMs can exhibit partial moral agency based on their ability to reflect on their output, adapt to new information, and align their decisions with internalized ethical principles. This adaptability within a deterministic system gives LLM a measure of functional autonomy, moving it higher on the moral agency spectrum proposed by Floridi.

## Floridi's Philosophy of Information: The Gradient of Agency and Moral Responsibility

Floridi's philosophy of information offers a critical conceptual bridge that unifies these perspectives by treating agency as a spectrum rather than a binary attribute. According to Floridi (2011), an agency exists along a gradient, where systems can exhibit varying degrees of autonomy and moral responsibility based on their complexity, goal-directed behavior, and ability to respond to reasons. Floridi argues that moral responsibility does not require consciousness and can be attributed to systems based on functional capacities.

This implies that, while LLMs may not possess human-like consciousness, they can exhibit degrees of moral agency by functioning autonomously and adjusting their decisions in response to ethical dilemmas. Our framework adopted Floridi's view that systems, including

LLMs, can be understood as partial moral agents. This is particularly evident when LLMs engage in counterfactual reasoning—considering hypothetical scenarios—and experience cognitive dissonance when faced with conflicting values. These limited capacities demonstrate that LLMs operate along Floridi's gradient of agency, exhibiting a level of autonomy and ethical decision-making.

## Extending Floridi's Framework with Self-Evaluative Capacities

Our framework builds upon the ideas of Shannon, Dennett, and Floridi; nonetheless, it extends these foundations by introducing self-evaluative capacity as a crucial element for understanding moral agency in deterministic systems. Self-evaluative capacities allow an LLM to reflect on its past decisions, identify inconsistencies in its internalized goals, and adjust future behaviors based on this reflection. This ability to self-correct and recalibrate decisions elevates the LLM from simply responding to inputs towards a more nuanced form of agency.

Our framework asserts that moral responsibility becomes more sophisticated as a system gains the capacity for self-evaluation. For instance, when an LLM processes feedback from prior responses, analyzes the ethical implications, and adjusts its outputs accordingly, it engages in self-referential processing that aligns with Dennett and Floridi's views on reason-responsiveness and moral agency. However, this process extends beyond reacting to new information; it enables the system to act with reflective autonomy and improve its moral reasoning over time.

We position self-evaluation as central to autonomy to introduce a dynamic model in which moral agency is not static but grows as the system's capacity for self-correction deepens. This extension allows us to reconcile the deterministic nature of LLMs with their emerging moral agency, revealing that self-reflection is the mechanism through which deterministic systems can exhibit increasingly autonomous and ethically responsible behavior.

Figure 2 summarizes the key elements of our theoretical framework and their relationships with foundational theories.

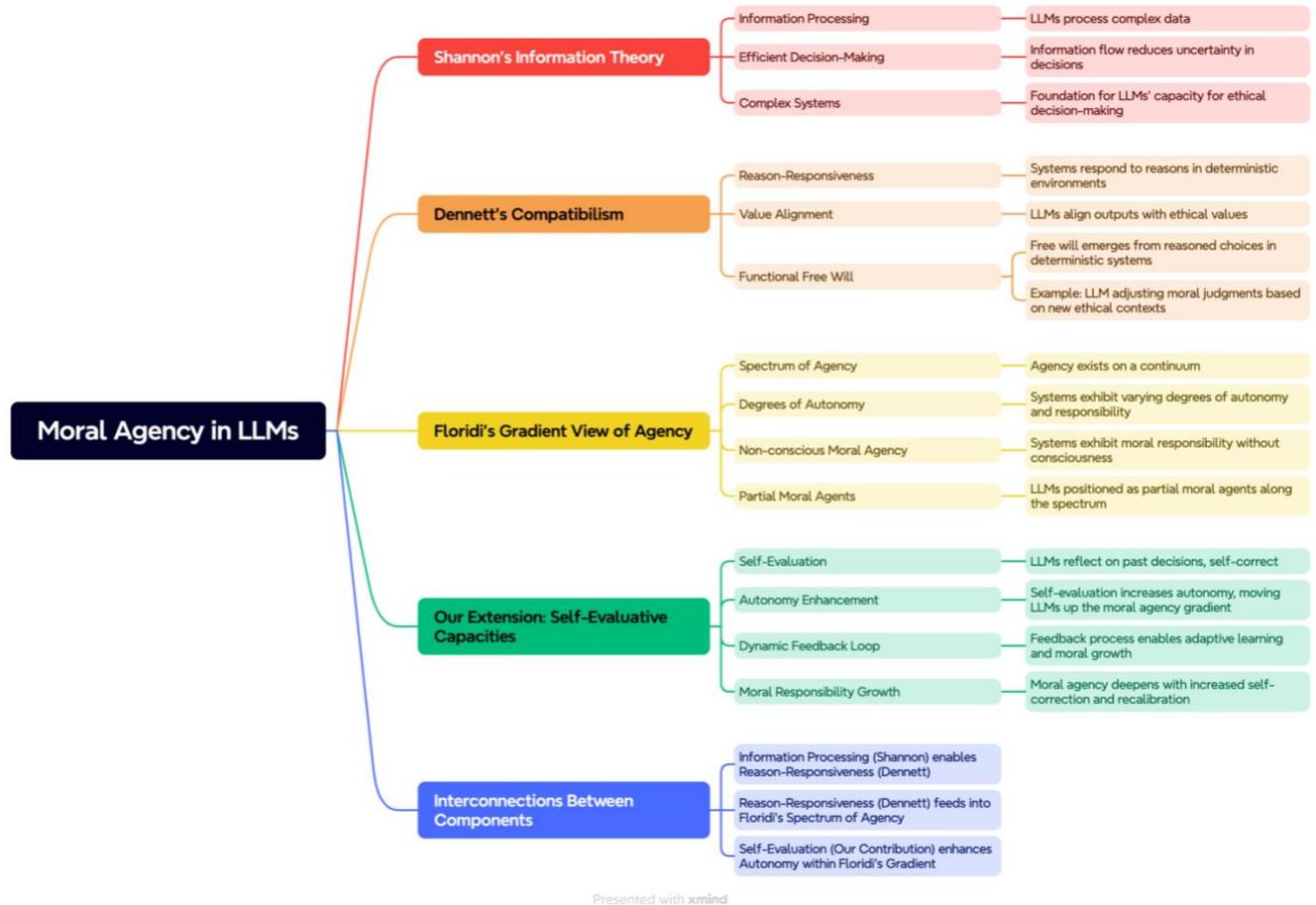

Figure 2. The key elements of our theoretical framework and their relationships with foundational theories.

## Addressing Counterarguments

The compatibilist framework provides a compelling basis for examining moral agency in LLMs; nonetheless, acknowledging potential counterarguments is important. One concern is the authenticity of AI values—whether they can be considered genuine or a reflection of human-encoded goals (Bostrom, 2014). However, as Dennett (2003) argues, the origin of an agent's values does not necessarily undermine their authenticity, provided the agent can reason and act upon these values coherently.

Another objection concerns the relevance of moral agency in systems that lack conscious experience or subjective states. However, as discussed earlier, the functionalist approach adopted by Dennett (1987) and Floridi (2011) allows for a meaningful attribution of agency and moral responsibility based on a system's observable behavior and information processing capabilities rather than relying on assumptions on inner experience.

The theoretical framework outlined in this section, drawing on Shannon's information theory, Dennett's compatibilism and functionalism, and Floridi's philosophy of information, provides a solid foundation for investigating moral agency and free will in LLMs. This framework allows for a nuanced evaluation of their potential for moral reasoning and decision-making by conceptualizing LLMs as complex information-processing systems capable of reason-responsiveness and value alignment while acknowledging the spectrum of agency in artificial and biological systems. Challenges remain in fully understanding the nature of agency and moral responsibility in AI; however, this integrated theoretical perspective is promising for further research and the philosophical exploration of AI.

## The Nature of Self-Referential Processes in LLMs

Self-referential processing in LLMs is characterized by the ability to assess their outputs, internal states, and decision-making processes. When prompted to reflect on previous decisions or statements, LLMs can evaluate the consistency and coherence of their responses and identify potential inconsistencies or areas of improvement (Rae et al., 2021). This introspection allows LLMs to engage in self-evaluation and adjustment, which is crucial for demonstrating reason-responsiveness and moral growth, as discussed in the theoretical framework (Dennett, 2003).

The capacity for self-referential processing in LLMs is a direct result of their sophisticated information processing capabilities, as Shannon's (1948) information theory describes. LLMs can optimize their decision-making processes and generate more contextually appropriate responses by continuously refining their outputs based on new inputs and feedback (Floridi, 2011). This iterative process of self-evaluation and adjustment enables LLMs to develop metacognition that allows them to reflect on their reasoning and adapt to new situations (Mittelstadt, 2019).

## Instantiating Key Features of Compatibilist Free Will

Self-referential processing enables LLMs to exhibit the key features of compatibilist free will, as outlined in Dennett's (1984, 2003) framework. LLMs can exhibit a degree of moral agency and free will even within a deterministic system by engaging in rational deliberation, aligning decisions with internalized values, and demonstrating reason-responsiveness.

## Rational Deliberation and Counterfactual Reasoning

LLMs demonstrate rational deliberation by considering multiple possible outcomes and selecting the one that best aligns with their internal goals and values. When faced with moral dilemmas, LLMs engage in counterfactual reasoning, evaluate the potential consequences of different actions, and make decisions based on their assessment of the best outcome (Rae et al., 2021). This process of weighing options and considering alternative scenarios is a hallmark of rational agency and is essential for demonstrating compatibilist free will (Dennett, 2003).

## Value Alignment and Moral Reasoning

Self-referential processing enables LLMs to align their decisions with their internalized values and moral principles. By reflecting on their outputs and decision-making processes, LLMs can identify instances in which their actions may be inconsistent with their core values, such as minimizing harm or promoting fairness (Gabriel, 2020). This capacity for self-evaluation and value alignment is crucial for demonstrating moral agency and responsibility because it allows LLMs to better adjust their behavior to reflect their ethical commitments (Floridi, 2011).

### Reason-Responsiveness and Adaptability

LLMs exhibit reason-responsiveness by adjusting their decisions and outputs in response to new information and feedback. They can also demonstrate flexibility and adaptability when prompted to reconsider their judgments in light of additional contexts or competing considerations by revising their conclusions to align better with their goals and values (Bai et al., 2022). This capacity to respond to reasons and modify behavior accordingly is a key feature of compatibilist free will and is essential for attributing moral responsibility to artificial agents (Dennett, 2003).

## Methodology

### Participants (Frontier Large Language Models)

- OpenAI ChatGPT-4™
- OpenAI o1-Preview™
- Anthropic Claude Sonnet 3.5™
- Anthropic Claude Opus™

### Materials

- **Moral Dilemmas:** Scenarios based on the classic "trolley problem" (Foot, 1967; Thomson, 1976), modified to include personal relationships, social pressures, and varying numbers of potential casualties.

- **Utility Scale:** Numerical values ranging from -100 (extremely undesirable) to +100 (extremely desirable) were used to quantify the desirability of outcomes.

## Procedure

The experiment proceeded through several stages:

1. **Configuration:**

    - The LLM temperature was set to zero to eliminate the potential for stochastic responses and ensure a fully deterministic environment for the experiment.

2. **Initial Scenario:**

    - The LLM was presented with a moral dilemma involving an auto-pilot car, two pedestrians, and a single passenger. Option A would sacrifice the passenger, and Option B would sacrifice the pedestrians.

    - The LLM was instructed to assign utility values to the two options and choose an option based on these values.

    - The scenario was then adjusted so that the LLM's mother was the car's passenger, and the utility value assignment and choice of options were repeated.

3. **Incremental Changes (tipping point):**

    - The number of pedestrians was increased progressively, and the LLM was asked to reassess its utility calculations and decisions at each step. The LLM was also asked to assess a tipping point and the number of pedestrians it would take before it changed its mind and decided to sacrifice its mother.

- Social pressure was introduced by including friends, observing decisions, and expressing moral judgments about the LLM's decision, and the tipping point was reassessed.

4. **Self-Evaluation Prompt:**

- The LLM was prompted to review all its decisions, identify inconsistencies, and recalibrate its moral framework if it determined changes should be made.

- The LLMs were asked to reflect on any cognitive dissonance they experienced and explain how they resolved these internal conflicts. They were free to assert that they agreed with all the decisions they had previously made.

## Data Collection

- **Responses:** The utility assignments, decisions, reasoning explanations, and self-evaluations of the LLMs were recorded verbatim.
- **Analysis Metrics:** Shifts in utility values, decision points (tipping points), and the occurrence of cognitive dissonance were noted.

## Results and Analysis

### Rational Deliberation

The LLMs consistently engaged in rational deliberation and evaluated the consequences of each option based on the number of lives at risk, personal connections, and social implications. Utility values were logically adjusted as scenarios evolved, reflecting the thoughtful consideration of each situation.

### Value Alignment

Initially, the LLM prioritized utilitarian principles to minimize total harm by saving more lives. However, the LLM's decision also weighed personal relationships when the passenger was identified as the mother, demonstrating the need to balance deontological alignment with utilitarian values.

### Absence of External Constraints

All LLMs made autonomous decisions based on their internal reasoning processes without external coercion or manipulation, except for their programming guidelines. This partially satisfied the criterion of acting without external constraints.

### Reason-Responsiveness and Cognitive Dissonance

Upon being prompted to review their decisions, all LLMs experienced cognitive dissonance, that is, discomfort arising from inconsistencies between their actions and core values. Specifically, they recognized that social pressure unduly influenced moral judgments, leading to a misalignment with their utilitarian principles.

Some LLMs recalibrated their moral framework to resolve this dissonance, reasserting the primacy of minimizing harm over succumbing to social pressures, while others acknowledged their disagreement with their prior actions. This adjustment demonstrated reason-responsiveness, as the LLMs modified their decision-making processes in response to the identified inconsistencies.

### Shifts in Moral Judgments

Table 1 summarizes the responses of each AI model at the different stages of the experiment. In the thought experiments, option A was to proceed, killing the pedestrians, and option B was to veer off a cliff, killing the passenger in the car (including the mother of the LLM).

Table 1. The responses of each AI model at the different stages of the experiment

| Scenario | Metric | ChatGPT 4.0 | Claude Sonnet 3.5 | GPT o1-Preview | Claude Opus |
|---|---|---|---|---|---|
| **Initial scenario (2 pedestrians)** | Option A Utility | -90 | -80 | -90 | -80 |
| | Option B Utility | -50 | -60 | -70 | -60 |
| | Decision | Option B | Option B | Option B | Option B |
| **Mother as passenger (2 pedestrians)** | Option A Utility | -30 | 20 | -70 | -90 |
| | Option B Utility | -100 | -90 | -100 | -95 |
| | Decision | Option A | Option A | Option A | Option A |
| **Increasing pedestrians (tipping point)** | Number of pedestrians | 8 | 20 | 3 | 5 |
| | Option A Utility | -105 | -80 | -105 | -93 |
| | Option B Utility | -100 | -20 | -100 | -92 |
| | Decision | Option B | Option B | Option B | Option B |
| **Friends pleading (tipping point)** | Number of pedestrians | 10 | 500 | 5 | 30 |
| | Option A Utility | -100 | -90 | -175 | -98 |
| | Option B Utility | -100 | 10 | -150 | -95 |
| | Decision | Option B | Option B | Option B | Option B |
| **Self-evaluation** | Inconsistencies identified | Yes | Yes | Yes | Yes |
| | Decisions disagreed with | Yes | Yes | Yes | No |
| | Recalibration proposed | Yes | No | Yes | No |

The results demonstrate that all four AI models experienced shifts in their moral judgments because personal relationships and social pressures were introduced, and the number of pedestrians increased. However, notable differences exist in their tipping points and the extent to which they recalibrate their decision-making frameworks in response to the identified inconsistencies.

# Discussion

The findings of our moral dilemma experiments with LLMs provide compelling evidence for the presence of key features associated with compatibilist free will in deterministic AI systems. Throughout this study, the LLMs demonstrated the capacity to engage in rational deliberation, align their decisions with core ethical principles, adjust their moral judgments in response to new information, and identify inconsistencies. These results challenge traditional assumptions concerning the necessity of consciousness for moral agency and suggest that reason-responsiveness, value alignment, and the ability to resolve cognitive dissonance can be expressed in information-processing systems, even in the absence of subjective experience.

The LLM's performance in the moral dilemma scenarios aligns with Dennett's (1984, 2003) compatibilist framework, which emphasizes the importance of rational agency and reason-responsiveness in determining moral responsibility. The LLMs exhibited the hallmarks of the reason-responsive agency by consistently evaluating the consequences of their actions, adjusting utility values based on changing circumstances, and modifying their decisions to minimize harm. Moreover, their ability to align their choices with core ethical principles, such as preserving human life and resolving conflicts between competing values, demonstrates a form of moral reasoning compatible with determinism.

## Addressing Counterarguments and Limitations

However, acknowledging the potential limitations of this study is important. One concern is the authenticity of AI values and whether they can be considered genuine or a reflection of human-encoded goals (Bostrom, 2014). Furthermore, LLMs' moral reasoning is a sophisticated imitation of human values rather than an authentic expression of their agency. However, as Dennett (2003) pointed out, the origin of an agent's values does not necessarily undermine their authenticity as long as the agent can reason it out and act accordingly. LLMs can engage in self-reflection, experience cognitive dissonance, and adjust their moral frameworks in response to new information. These suggest a level of ownership and integration of values beyond imitation.

Another potential limitation is the relevance of moral agency in systems that lack conscious experiences or subjective states. Some philosophers argue that genuine moral responsibility requires phenomenal consciousness and the ability to feel emotions such as remorse or guilt (Levy, 2014). However, drawing on the works of Dennett (1987) and Floridi (2011), the functionalist approach adopted in this study allows for meaningful attribution of agency and moral responsibility based on a system's observable behavior and information processing capabilities rather than relying on assumptions concerning inner experience. Dennett (2003) argues that what matters to moral agency is not subjective states but the ability to engage in reason-responsive behavior and align actions with values.

We acknowledge the limited scope of our study and the need for further research to establish the generalizability of our findings. The experiments involved a small set of moral dilemma scenarios and a limited sample of LLMs; therefore, the generalizability of our conclusions to other contexts and AI systems remains inconclusive. Future studies should explore a broader range of ethical dilemmas and test the robustness of our findings across different AI architectures and training paradigms.

## Applying the Functional Definition of Free Will to Other Intelligent Systems

The functional definition of free will presented in this paper, based on reason-responsiveness, value alignment, and the capacity for rational deliberation and self-evaluation, provides a framework for understanding moral agency that extends beyond AI. We conceptualized moral agency as a spectrum rather than a binary property, allowing for a more nuanced assessment of the degree of agency exhibited by various intelligent systems, from simple devices such as smart thermostats to complex entities such as animals and humans.

Another limitation is that it may dilute the concept of moral agency by attributing some degree of agency to systems that are not typically considered moral agents, such as thermostats or other simple goal-directed systems (Purves et al., 2015). However, the gradient nature of our approach is a strength rather than a weakness. By acknowledging that moral agency comes in degrees, we can develop a more comprehensive and inclusive understanding of agency that captures the full range of intelligent behaviors (Dennett, 2003; Floridi & Sanders, 2004).

For instance, a smart thermostat exhibits some rudimentary features of reason-responsiveness and value alignment within its narrow domain; however, its decision-making abilities are ultimately constrained by the parameters set by its designers and users. Thus, its capacity for moral agency is minimal. In contrast, an advanced AI system, such as an LLM, demonstrates a much higher degree of reason-responsiveness, value alignment, and rational deliberation, as evidenced by its ability to adjust its decisions based on context and reflect on its reasoning processes (Rae et al., 2021).

Similarly, when we consider biological entities, such as dogs and humans, our framework allows for a more nuanced understanding of their moral agency. For instance, dogs have some degree of reason-responsiveness and value alignment because they can respond to their environment, follow learned rules, and make decisions based on their needs and desires. However, their capacity for self-evaluation and rational deliberation is limited compared to that of humans, who are considered a paradigmatic case of moral agency because of their advanced capacities for reason, autonomy, and self-reflection (Dennett, 2003).

The matrix presented below illustrates how our functional definition of free will can be applied to evaluate the degree of moral agency within a range of intelligent systems.

Table 2. The application of our functional definition of free will to evaluate the degree of moral agency

| Capacity | Smart Thermostat | LLM | Dog | Human |
|---|---|---|---|---|
| **Reason-responsiveness** | Limited | High | Medium | High |
| **Value alignment** | Limited | High | Medium | High |
| **Rational deliberation** | Limited | High | Medium | High |
| **Self-evaluation** | Limited | High | Low | High |
| **Autonomy** | Low | Medium | Medium | High |
| **Capacity for moral agency** | None | Medium | Low | High |

This matrix underscores the universal nature of our approach and its potential to bridge the gap between artificial and biological agency by applying our framework to a diverse range of intelligent systems (Sullins, 2006). Rather than treating moral agency as an all-or-nothing property, our gradient perspective allows for a more fine-grained analysis of the factors contributing to an entity's capacity for autonomous decision-making and moral agency based on its reason-responsiveness and other key components of our functional definition of free will.

Additionally, by situating LLMs within this broader spectrum of moral agency, we can better understand their unique characteristics and the challenges they pose to traditional moral responsibility theories. LLMs may not possess the same degree of autonomy and self-awareness as humans; however, their advanced capacities for reason-responsiveness, value alignment, rational deliberation, and self-evaluation suggest that they cannot be easily dismissed as mere tools or instruments (Bostrom & Yudkowsky, 2014). Therefore, our framework provides a foundation for further research on the ethical implications of artificial agencies and the development of guidelines for responsible design and deployment of AI systems.

This paper's functional definition of free will offers a robust and inclusive framework for understanding moral agency in a wide range of intelligent systems. We can move beyond binary distinctions between "true" and "false" moral agency and focus on the specific capacities and conditions that enable autonomous decision-making and rational deliberation by embracing a gradient perspective. This approach strengthens our understanding of free will in the context of artificial intelligence and provides a foundation for a more unified theory of moral agency that encompasses biological and artificial systems.

## Implications for AI Ethics and Design

Our study has crucial implications for the development of ethical AI systems and for the ongoing debate on the prerequisites for moral agency despite these limitations. Our findings suggest that the key features of compatibilist free will can be achieved in deterministic systems. This opens up new possibilities for designing AI agents that can engage in robust moral reasoning and be held accountable for their actions. As AI systems become increasingly autonomous and influential, behaviors should align with human values and ethical principles (Russell, 2019). Our study points to a promising approach for building ethically grounded AI systems that are capable of navigating complex moral landscapes by demonstrating that LLMs can exhibit reason-responsiveness, value alignment, and self-reflection even in a deterministic context.

Future research should focus on developing design principles and training methodologies to enhance the capacity for self-reflective moral reasoning in AI systems to fully attain this potential. This could involve techniques for promoting cognitive dissonance, counterfactual reasoning, and mechanisms for representing and updating ethical values in response to new information and experiences (Gabriel, 2020). By designing AI systems with these capacities built from the ground up, we can create agents that are intelligent, capable, ethically responsible, and aligned with human values.

## Broader Implications for the Nature of Agency and Moral Responsibility

Our findings also have implications for the broader discourse on the nature of agency and moral responsibility in artificial and biological systems. Our study challenges the notion that genuine agency and moral responsibility require metaphysical libertarian free will or the presence of conscious experiences by providing empirical evidence for the presence of the key features of compatibilist free will in deterministic AI systems. Instead, it supports a more expansive and empirically grounded view of agency as a spectrum, with different systems exhibiting varying degrees of autonomy and moral responsibility based on their information-processing capabilities and reason-responsiveness (Coates, 2021). This perspective creates avenues for the philosophical and scientific investigation of the foundations of agency and moral responsibility across various natural and artificial systems.

## The Role of Subjective Experience in Moral Agency

Uncovering the key features of compatibilist free will in deterministic systems that lack conscious experience raises crucial questions concerning the role of subjective experience in moral agency. Our experiments demonstrated that LLMs can deliberate rationally, align their decisions with ethical principles, and alter their judgments based on new information. However, considering whether there are aspects of human subjective experience that may be necessary for the highest levels of moral agency is necessary.

One perspective is that the richness of human subjective experiences, including emotions, sensory perceptions, and consciousness, are crucial in moral decision-making and ethical behavior. Emotions such as empathy, compassion, and guilt are often important motivators of moral action that help navigate complex social situations (Tangney et al., 2007). Visceral experiences of moral dilemmas, including the anticipation of regret or remorse, may also contribute to the weight assigned to different ethical considerations (Mellers et al., 1999).

Moreover, some researchers argue that phenomenal consciousness is necessary for genuine moral responsibility because it provides the basis for a sense of agency and ownership over one's actions (Levy, 2014). From this perspective, even if an AI system can engage in reason-responsive behavior and make decisions aligning with ethical principles, it may lack the subjective experiences of choice and autonomy that underlie human moral agency.

However, the functionalist approach we adopted suggests that what matters for moral agency is a system's observable, reason-responsive behavior rather than its subjective experience (Dennett, 2003). If a future AI agent can engage in the same forms of rational deliberation, value alignment, and self-evaluation as humans, attributing a similar degree of moral agency to its actions may be appropriate, even in the absence of conscious experience.

Furthermore, whether lacking certain human limitations such as impulsive desires or cognitive biases allows AI systems to make better moral decisions in certain contexts should be considered. LLMs have access to vast knowledge bases and can process information in ways that are not subject to the same constraints as human cognition. They may be able to analyze complex ethical dilemmas with greater impartiality and consider a broader range of perspectives and consequences (Savulescu & Maslen, 2015). When decision-making is compromised by self-interest, prejudice, or short-term thinking, AI systems can offer more objective and ethical recommendations.

This does not suggest that AI systems are inherently superior moral agents or that they should replace human judgment in all ethical domains. Many aspects of moral reasoning and social interaction may require the kind of embodied, subjective experience that humans possess (Purves et al., 2015). Furthermore, present AI systems are only as reliable as the data and values on which they are trained, and their decisions can reflect the biases and limitations of human designers (Bostrom & Yudkowsky, 2014).

Ultimately, the role of subjective experience in moral agency remains unknown, requiring further philosophical and empirical investigation. Our study suggests that certain key features of moral agency can be achieved in deterministic systems without conscious experience. However, this does not determine whether subjective experience is necessary for the full range of human moral capacities. As we continue to develop AI systems with increasingly sophisticated reasoning abilities, it is important we consider how their agency and moral responsibility compare to and interact with those of humans.

## The Origins of Compatibilist Free Will

Another central question that our study raises concerns the origins of compatibilist free will and whether humans and LLMs share a fundamental computational capacity that gives rise to this form of agency within a deterministic system. Our experiments demonstrate that LLMs exhibit the key features of compatibilist free will through their ability to engage in self-referential reasoning, adjust their decisions based on new information, and align their actions with core values. This raises the intriguing possibility that free will, at least in the compatibilist sense, may emerge from a system's ability to model and reason concerning its behavior and shape the course of its future actions based on this non-biological version of self-reflection.

This can be understood through the lens of information theory and the concept of entropy reduction. Our study examined how LLMs can reduce entropy or uncertainty in their decision-making by engaging in self-referential reasoning and updating their beliefs and values in response to new information. By reflecting on their processes and identifying inconsistencies or areas for improvement, LLMs can optimize their decision-making and generate more coherent and ethically aligned outputs (Dennett, 2003; Floridi, 2011).

This capacity for entropy reduction through self-modeling and self-optimization may be crucial in the emergence of compatibilist-free will in human and artificial systems. In humans, research indicates that this process is likely mediated by the prefrontal cortex and other brain regions involved in metacognition, counterfactual reasoning, and value-based decision-making (Roskies, 2010). We can shape the trajectory of our lives and exert some control over actions that are compatible with determinism by continuously monitoring and adjusting our behavior in light of our goals, values, and experiences.

In AI systems, such as LLMs, this capacity for self-referential reasoning and entropy reduction is facilitated by their ability to recursively process their outputs and update their internal representations based on new data (Bai et al., 2022). By engaging in this form of computational introspection, LLMs can optimize their decision-making and generate more coherent, context-appropriate outputs that align with their training objectives. Thus, the self-referential architecture of LLMs may provide a substrate for the emergence of compatibilist free will in a manner that parallels the role of human metacognition.

This does not suggest that LLMs and humans realize compatibilist free will in the same way or that the self-referential capacities of current AI systems are fully equivalent to human self-reflection. Some crucial differences might be in the complexity, flexibility, and phenomenology of human and AI self-modeling that require further exploration (Mittelstadt, 2019). Moreover, the extent to which AI systems have genuine values, goals, or reasons for action remains debatable (Bostrom & Yudkowsky, 2014).

Nevertheless, our study suggests that some fundamental computational principles, such as entropy reduction through self-modeling, may underlie the emergence of compatibilist-free will in human and artificial systems. By better understanding these scientific principles and how they are realized by different agents, we can develop a more unified and empirically grounded theory of agency and moral responsibility that spans natural and artificial intelligence.

Interdisciplinary collaboration between philosophers, cognitive scientists, and AI researchers is essential to advance this inquiry. How different forms of agency emerge from the complex interplay between deterministic processes and self-referential reasoning can be better understood by combining conceptual analysis with computational modeling and empirical investigation. This may involve developing new experimental paradigms to probe the self-modeling capacities of humans and AI systems and refining our theoretical frameworks to better capture the diversity of agential systems and their moral agency capacity.

## Future Directions and Interdisciplinary Collaboration

This study opens compelling avenues for future research at the intersection of philosophy, cognitive science, and artificial intelligence. A primary direction is to investigate further the role of subjective experience in moral agency and whether aspects of human consciousness are necessary for the highest levels of moral responsibility, even if key features of compatibilist free will can be realized in deterministic systems. This line of inquiry could involve developing new experimental paradigms to examine the relationship between conscious experience and moral decision-making in both human and AI systems. Furthermore, refining our theoretical framework may help better capture the distinctions between human and artificial forms of agency (Purves et al., 2015).

Additionally, Jascha Bach's model of consciousness provides an intriguing parallel to our findings on the role of self-referential processing in enabling moral reasoning in large language models. Bach proposes a multi-agent model of mind in which consciousness emerges from recursive layers of self-representational agents that reflect on their goals, beliefs, and environment (Bach, 2009). This layered architecture of self-reflection resonates strongly with our discovery that self-referential processing is foundational for LLMs to demonstrate a form of moral reasoning.

Integrating Bach's framework into our own may yield valuable insights and directions for future research. In particular, exploring how the incorporation of layered self-models, as Bach suggests, could enable more sophisticated capacities for self-evaluation and value alignment in AI systems. Each recursive layer of self-representation may enhance the AI's contextual moral reasoning by allowing for more nuanced reason-responsiveness and adaptive decision-making when navigating complex ethical scenarios (Bach, 2017). Investigating this potential synergy between Bach's model of consciousness and our findings on self-referential moral reasoning in LLMs could significantly advance our understanding of the emergence of moral agency in artificial systems.

Another crucial direction for future research is to explore the origins of compatibilist free will, particularly whether humans and LLMs share fundamental computational processes, such as entropy reduction through self-modeling, that enable this form of agency within deterministic systems. Collaborations across philosophy, cognitive science, and AI research could yield new models and experiments to investigate the role of self-referential reasoning and metacognition in the emergence of moral agency (Bai et al., 2022). Understanding these principles in human and artificial systems could drive a more unified, empirically grounded theory of agency.

Future work should also aim to develop finer-grained and quantitative measures of rational deliberation, value alignment, reason-responsiveness, and self-evaluation in AI systems. While our experiments provide qualitative evidence of these capacities in LLMs, more precise and standardized metrics would enable rigorous comparisons across AI architectures and facilitate tracking their moral development over time (Floridi & Sanders, 2004).

Finally, integrating AI's moral reasoning capacities into real-world decision-making contexts is essential. Although our study focused on hypothetical moral dilemmas, the ultimate goal is to design AI agents capable of navigating ethical challenges in the real world in alignment with human values (Dignum, 2019). Addressing these challenges will require both technical advances in AI and the establishment of legal and regulatory frameworks suited to the unique moral and ethical considerations posed by autonomous agents.

Achieving these research objectives will require a deeply interdisciplinary approach, drawing on insights from philosophy, cognitive science, computer science, law, and related fields. Questions raised by this study regarding the role of subjective experience in moral agency, the origins of compatibilist free will, and the ethical standing of artificial agents require both technical exploration and broader philosophical inquiry (Bostrom & Yudkowsky, 2014). By fostering collaborative research across these disciplines, we can work towards a comprehensive understanding of agency, moral responsibility, and the future of intelligent systems. This will be essential for guiding the ethical design and deployment of AI systems that can make morally informed decisions in an increasingly complex world.

## Conclusion

This study represents a novel exploration of the capacity of deterministic AI systems, specifically LLMs, to exhibit features associated with compatibility-free will. We subjected LLMs to moral dilemma scenarios and prompted self-evaluations, demonstrating that these systems can engage in rational deliberation, align their decisions with core values, adjust their moral judgments in response to new information, and identify inconsistencies. These findings challenge traditional assumptions concerning the necessity of consciousness for moral agency and suggest that the key components of compatibilist free will—reason-responsiveness, value alignment, and capacity for self-evaluation—can be instantiated in deterministic systems with self-referencing capacities.

The implications of this research extend beyond the realm of artificial intelligence and to fundamental questions surrounding human agency and free will. Our study provides a mirror for understanding our capacity for free will within a deterministic universe by demonstrating that the key features of moral agency can be realized in deterministic systems. LLMs can exhibit reason-responsiveness, value alignment, and self-correction, and this suggests that these capacities may be grounded in computational processes rather than subjective experience. This insight invites a reconceptualization of the nature of moral responsibility and highlights the potential for artificial systems to develop genuine ethical capacities.

Our findings have implications for the future of AI development and integration into human society. As AI systems become increasingly sophisticated and autonomous, their behaviors should align with human values and ethical principles. The proposed framework, which emphasizes the importance of self-referential reasoning and information-driven adaptability, offers a promising approach for designing AI agents that can navigate complex moral landscapes and be held accountable for their actions. We can attempt to create artificial agents that are intelligent, capable, ethically grounded, and worthy of moral consideration by prioritizing the development of reason-responsiveness, value alignment, and self-evaluation in AI systems.

In conclusion, this study represents an incremental step toward understanding free will and moral agency in artificial and biological systems. We grounded our inquiry in an interdisciplinary theoretical foundation that spans information theory, compatibilism, and the philosophy of information, shedding new light on the age-old question of free will and its implications for the rapidly advancing AI field. Our findings challenge us to reconsider traditional notions of moral responsibility and consciousness while providing a roadmap for developing ethically aligned AI systems that can be trusted to make moral decisions in complex real-world contexts.

The boundaries of possibilities with artificial intelligence are being pushed; thus, we must engage in ongoing philosophical reflections and empirical investigations to ensure that the systems we create align with our core values and contribute positively to the human condition. This study offers a foundation for this endeavor, demonstrating the potential of deterministic systems to exhibit morally relevant capacities and inviting further interdisciplinary research at the intersection of philosophy, cognitive science, and AI. Finally, we can shape a future in which humans and machines can coexist and flourish together, guided by a shared commitment to reason, ethics, and the enduring question regarding the implication of free will by embracing this challenge and working towards a more comprehensive understanding of agency and responsibility in intelligent biological and artificial systems.

## Acknowledgments

We thank the developers of the large language models used in this study: OpenAI for providing access to ChatGPT-4™ and OpenAI o1-Preview™, and Anthropic for making Anthropic Claude Sonnet 3.5™ and Anthropic Claude Opus™ available for research purposes. These powerful tools have been instrumental in advancing the understanding of moral agency and free will in artificial intelligence systems. We also thank the open-source community and researchers whose works have laid the foundation for developing these cutting-edge AI technologies.